\documentclass[notitlepage, twoside,english,nofootinbib,preprintnumbers,aps,11pt,showpacs, superscriptaddress]{revtex4-1} \usepackage[a4paper, margin=1.5in]{geometry} \usepackage{amssymb} \usepackage{commath}
\usepackage{amsmath}
\usepackage{url}
\usepackage{float}
 \usepackage[utf8]{inputenc} \usepackage[T1]{fontenc} \usepackage{lmodern}
\usepackage{braket}
\usepackage{graphicx}
\usepackage{pictexwd,dcpic}
\usepackage{braket}
\usepackage{atbegshi,picture, afterpage}
\usepackage{lipsum}
\usepackage{physics}
\newcommand{\be}{\begin{equation}}
\newcommand{\ee}{\end{equation}}
\newcommand{\ba}{\begin{align}}
\newcommand{\ea}{\end{align}}
\usepackage{tkz-euclide}
\usetikzlibrary{arrows.meta}

\begin{document}
\title{Casimir effect in conformally flat spacetimes}
\author{Bartosz Markowicz}
\email[]{b.markowicz3@student.uw.edu.pl}
\address{\vspace{6pt} Institute of Theoretical Physics, Faculty of
  Physics, University of Warsaw, Pasteura 5, 02-093 Warsaw, Poland}
\author{Kacper Dębski}
\email[]{Kacper.Debski@fuw.edu.pl}
\address{\vspace{6pt} Institute of Theoretical Physics, Faculty of
  Physics, University of Warsaw, Pasteura 5, 02-093 Warsaw, Poland}
\author{Maciej Kolanowski}
\email[]{Maciej.Kolanowski@fuw.edu.pl}
\address{\vspace{6pt} Institute of Theoretical Physics, Faculty of
  Physics, University of Warsaw, Pasteura 5, 02-093 Warsaw, Poland}
\author{Wojciech Kamiński}
\email[]{Wojciech.Kaminski@fuw.edu.pl}
\affiliation{\vspace{6pt} Institute of Theoretical Physics, Faculty of
  Physics, University of Warsaw, Pasteura 5, 02-093 Warsaw, Poland}
  \author{Andrzej Dragan}
\email[]{Andrzej.Dragan@fuw.edu.pl}
\affiliation{\vspace{6pt} Institute of Theoretical Physics, Faculty of
  Physics, University of Warsaw, Pasteura 5, 02-093 Warsaw, Poland}
  \affiliation{\vspace{6pt} Centre for Quantum Technologies, National University of Singapore, 3 Science Drive 2, 117543 Singapore, Singapore}
\date{\today}
\begin{abstract}
We discuss several approaches to determine the Casimir force in inertial frames of reference in different dimensions.  On an example of a simple model involving mirrors in Rindler spacetime we show that Casimir's and Lifschitz's methods are inequivalent and only latter can be generalized to other spacetime geometries.  For conformally coupled fields  we derive the Casimir force in conformally flat spacetimes utilizing an anomaly and provide explicit examples in the Friedmann–Lemaître–Robertson–Walker $(k=0)$ models.
\end{abstract}
\maketitle
\section{Introduction}
Many years after the original work by Casimir  \cite{Casimir}, his effect is still considered bizarre. The force between two plates in a vacuum that was first derived using a formula originating from classical mechanics relates the pressure experienced by a plate to the gradient of the system's energy.
An alternative approach to this problem was introduced by Lifshitz \cite{Lifshitz:1956zz}, who expressed the Casimir force in terms of the stress energy tensor of the field. The force was written as a surface integral as for the static field  in the classical field theory. Both approaches are known to be equivalent in inertial scenarios \cite{dzyaloshinskii1959vanderwaals, dzyaloshinskii1961general}. Introducing non-inertial motion adds a new layer of complexity to the problem. Although Rindler metric that describes uniform acceleration is still stationary, the situation is more similar to a curved spacetime in many respects.

New phenomena arise when objects undergoing sudden acceleration emit radiation by the so-called dynamical Casimir effect \cite{Moore, Birrell:1982ix, Bruschi:2013}. However, cavity moving in an appropriate manner is still stationary system and thus we can ask what is a generalization of the Casimir force to this setting. This problem attracted much attention, with many authors using Casimir original approach. However, we will point out on a simple example, that in non-inertial frames the two definitions are not equivalent. We will also argue, that only Lifshitz approach is correct in this setting, pointing out that the source of the problem lies in the definition of the force as the derivative of the energy with respect to the position of the mirror.

Using a simple model of an accelerating cavity, we revise general methods of finding the Casimir force presented before. We compare results obtained in both approaches finding discrepancies. Surprisingly, energetic derivation leads to unphysical conclusions.
At the same time, it is a special case of the Casimir force evaluated in a conformally flat space-time. 
We also point out several misconceptions present in the literature and propose a possible generalization of our result beyond Rindler spacetime. Using conformal anomaly methods first formulated in \cite{Brown:1977sj}, we are able to derive the Casimir force in more general context. In particular, we calculated it in de Sitter background and presented all formulas needed in the general Friedmann--Lema\^itre--Robertson--Walker ($k=0$) universe. Our final results of Casimir forces in the Rindler and de Sitter spacetime have the potential of being used in some kind of an analog experiment testing quantum field theory in curved spacetime. We also comment on possible ambiguities that may arise in such setting.

The work is structured as follows. In sec. \ref{BM roz A} we present derivation of Casimir force in uniformly accelerating 1+1 dimensional cavity via Casimir Energy. This is the most popular approach to derivation of the Casimir effect \cite{Casimir, Avagyan:2002}. However, the Casimir Energy method in non-inertial reference frame leads to incorrect results, as we show in Section \ref{BM roz C}.  Sec.~\ref{BM roz B} provides correct derivation of Casimir force for a uniformly accelerating 1+1 dimensional cavity. The derivation presented in this chapter can be easily generalized for cavities moving along any trajectory in flat or conformally flat spacetimes. In sec. \ref{BM roz C} we compare the two described above methods of deriving Casimir force. We also propose how to fix a erroneous solution obtained via Casimir energy in case of mirrors with very simple geometry (ie. in the case that the passage of time at each point of the considered plate is the same).  Sec.~\ref{MK roz D} contains a discussion about the impact of the state outside the cavity on the observed Casimir effect.  In Sec.~\ref{MK roz III} we describe methods to derive Casimir force in any conformally flat space-time with the use of conformal anomaly we apply this approach to the case of  uniformly accelerating plates in 1+1 dimensional space-time in Sec.~\ref{MK roz IVA}. Finally, in  Sec.~\ref{MK roz IVB} we derive Casimir force for plates in 3+1 dimenional de Sitter spacetime.

\subsection{Conventions and notational remarks}
We follow sign conventions of \cite{Birrell:1982ix} and thus the metric signature is either $(+-)$ or $(+---)$ in two and four dimensions, respectively. Greek letters denote spacetime and Latin letters denote space indices. Riemann tensor is
\begin{equation}
    R^\alpha_{\ \beta \gamma \delta} = \partial_\delta \Gamma^{\alpha}_{\ \beta \gamma} -...
\end{equation}
while Ricci tensor is
\begin{equation}
    R_{\mu \nu} = R^\alpha_{\ \mu \alpha \nu}.
\end{equation}
Weyl tensor $C^\alpha_{\ \beta \gamma \delta}$ is a trace-less part of the Riemann tensor. \\
We are going to work both within a cavity with Dirichlet boundary conditions and in the whole spacetime. To distinguish between those cases, we will denote vacuum in the former case by $\ket{0}$ and in the latter by $\ket{\Theta}$. 
\section{Casimir Force in an accelerated frame}
\subsection{Setting \label{BM Set}} \label{setting}
\begin{figure}[h]
\begin{center}

\includegraphics[]{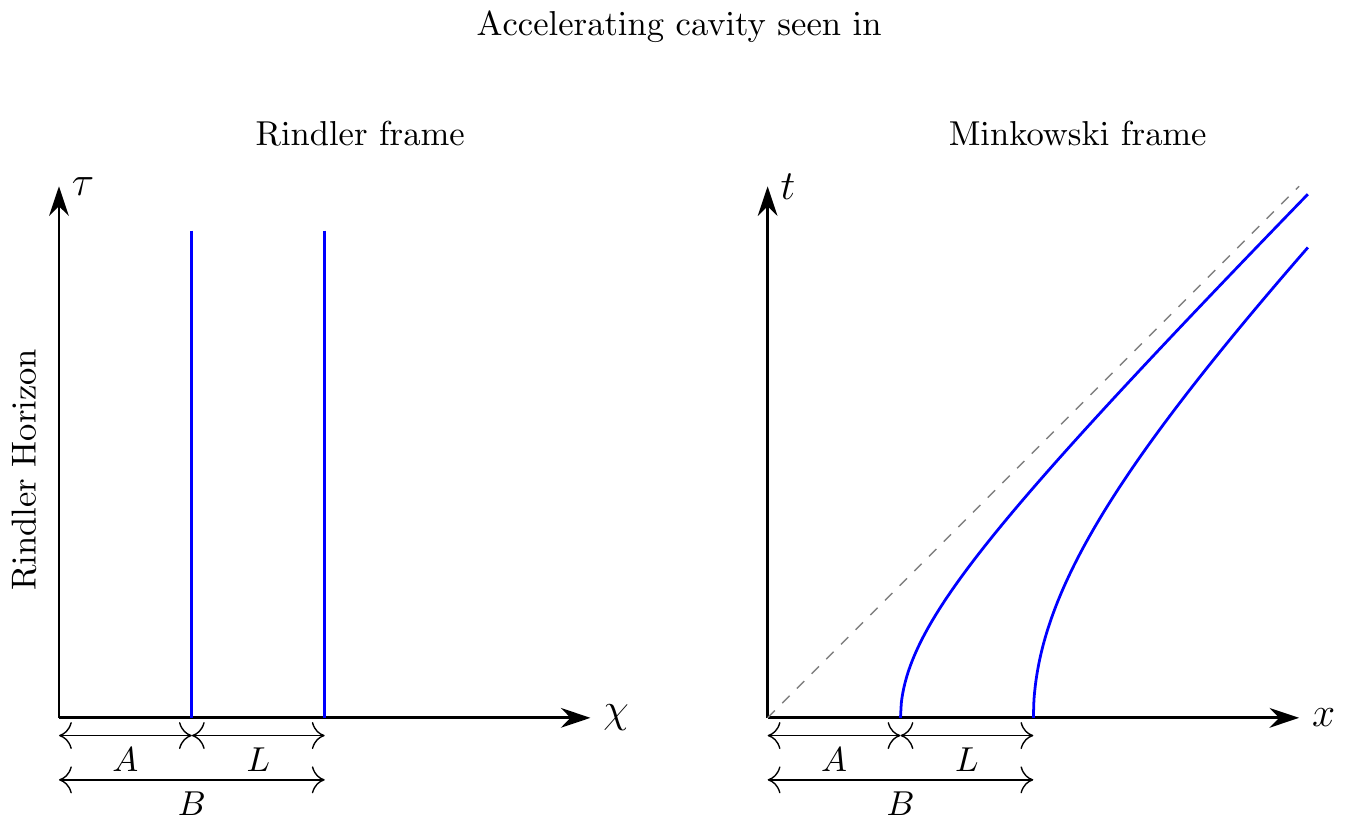}

\caption{The graphs show the space-time trajectories of uniformly accelerating plates in the Rindler frame $(\tau,\chi)$ and in the inertial Minkowski frame $(t,x)$. In the inertial frame at $ t = 0 $ both plates are in rest.}
\label{BM PIC}
\end{center}
\end{figure}

We consider a 1 + 1 dimensional cavity limited by uniformly accelerating plates (due to low dimensionality of our system, they are spatially point-like). They impose Dirichlet boundary conditions upon a scalar field in question.
The plates move in such a way that at any time in the cavity's rest frame the length of the cavity is constant and equal to $ L $.
To operate in the mentioned frame, we introduce Rindler coordinates $(\tau,\chi)$ associated with Minkowski coordinates $(t,x)$ through the following transformation 
\begin{equation}
    \chi=\sqrt{x^2-t^2} \;\;\;\;\;\;\;\;\;\;\;\;\;\; \tau=\frac{1}{a}\tanh^{-1} \frac{t}{x},\label{BM rindler}
\end{equation} 
where $a\in \mathbb{R}_+$ is a transformation parameter \cite{Rindler1969}. 
A physical interpretation of these coordinates is as follows.
A body moving on a trajectory $\chi=\textrm{const.}$ has a constant acceleration $\mathcal{A}_{\chi}=\frac{1}{\chi}$ and from (\ref{BM rindler}) it is easy to see that it follows a hyperbolic  trajectory from the point of view of an inertial observer \cite{Rindler1969} 
\begin{equation}
    x(t)=\sqrt{\frac{1}{\mathcal{A}_{\chi}^2}+t^2} \label{BM Rindler Placement}
\end{equation} with velocity 
\begin{equation}
    v(t)=\frac{\mathcal{A}_{\chi}t}{\sqrt{1+\mathcal{A}_{\chi}^2t^2}}. \label{BM Rindler Velocity}
\end{equation}
Metric tensor in the Rindler coordinates takes the form
\begin{equation}
    \dd s^2=(a\chi)^2 \dd\tau^2-\dd\chi^2.
\end{equation}
The plates are located at $ \chi=A $ and $ \chi=B = A + L $, as shown in the Fig. \ref{BM PIC}.

The normalized solutions to the wave equation in the Rindler coordinates take the form
\begin{equation}
    \psi(k)=\frac{1}{\sqrt{4\pi|k|}}e^{i\frac{k}{a}\log a \chi-i|k|\tau}, \label{BM mody ciagle}
\end{equation}
where $k \in\mathbb{R}$. 
With such a mode decomposition, we can associate a field operator
\begin{equation}
    \hat{\psi}=\int_{-\infty}^{\infty}\dd k \psi(k) \hat{a}_k + h.c., \label{BM mody ciagle 2}
\end{equation}
Operators $\hat{a}_k$ and $\hat{a}_k^{\dagger}$ in (\ref{BM mody ciagle 2}) are bosonic annihilation and creation operators, respectively and they satisfy the canonical commutation relation $\comm{\hat{a}_k}{\hat{a}_l^\dagger}=2\pi \delta(k-l)$. 
We define vacuum state as $\ket{\Theta}$ such that $\forall_k ~\hat{a}_k\ket{\Theta}=0$. \\ Solutions considered so far were obtained in an empty spacetime. However, we are mainly interested in the field theory within a cavity. Thus, we need to find modes which satisfy Dirichlet boundary conditions. They are of the form
\begin{equation}
    \psi_n(t,x)=\frac{1}{\sqrt{n \pi}} 
    \sin
    \left(
    n\pi\frac{\log\frac{\chi}{A}}{\log\frac{B}{A}}
    \right) e^{-i\frac{a n\pi}{\log \frac{B}{A}} \tau}, \label{BM mody dyskretne}
\end{equation}
where $ n $ is a positive integer. With such a mode decomposition of the field we can associate a field operator
\begin{equation}
    \hat{\psi}=\sum_{n=1}^{\infty} \psi_n \hat{b}_n + h.c., \label{BM mody dyskretne 2}
\end{equation}
where $\comm{\hat{b}_k}{\hat{b}_l^\dagger}=\delta_{kl}$. 
And now, (cavity) vacuum state is denoted as $\ket{0}$ such that $\forall_k~\hat{b}_k\ket{0}=0$.
Casimir force can be defined in two different ways. First as the work needed to move one of the plates by an unit distance. Intuitively, it will increase the energy of the field between the plates. The second definition is the pressure that the field exerts on the plates limiting the cavity. In this chapter, we examine both approaches and analyze the results they lead to.

\subsection{Casimir force as a work derivative \label{BM roz A}}
In 1948 Casimir in his original work \cite{Casimir}  derived force acting on the infinite parallel plates resting  in Minkowski spacetime. 
Since that time, many methods of determining Casimir's force have been proposed, however the original approach, thanks to its simplicity, remained the most popular.
Here we present an analogous derivation for accelerating plates.

In this approach we want to calculate the Casimir force in the accelerating frame we must first determine the difference between energy of the field located in the cavity and energy of the free field located in the area $ \chi \in [A, B] $. This difference will be referred as Casimir energy. Force can be then defined as a variation of that energy upon a small translation of a mirror.

The canonical energy of the field associated with the Killing vector $n^\mu \partial_\mu = \partial_\tau$ is defined as \cite{Wald:1984rg}
\begin{equation}
E_{AB}=\int_A^{B}\dd\chi\sqrt{|\det h|}\bra{0}\hat{T}_{\mu\nu}n^{\mu}v^{\nu}\ket{0}, \label{BM energy}
\end{equation}
where $v^{\nu}\partial_\nu=\frac{1}{a\chi}\partial\tau$ is a unit timelike vector orthogonal to the Cauchy surface $\tau = \textrm{const.}$, $h=-\mathrm{d}\chi^2$ is metric induced on this surface while $T_{\mu\nu}$ is a canonical stress--energy tensor of Klein--Gordon field and $\hat{T}_{\mu \nu}$ is associated quantum operator.
\eqref{BM energy} reduces itself into a simpler form
\begin{equation}
E_{AB}=\int_A^{B}\frac{1}{a\chi}\bra{0}\hat{T}_{\tau \tau}\ket{0}\dd\chi. \label{BM energy}
\end{equation}

Using the expression for the stress energy tensor applied to massless Klein-Gordon field \cite{Wald:1984rg}
\begin{equation}
        T_{\mu\nu}[\psi]=\partial_{\mu}\psi\partial_{\nu}\psi-\frac{1}{2}g_{\mu\nu}g^{\rho\xi } \partial_{\rho}\psi  \partial_{\xi}\psi, \label{BM ep tensor}
\end{equation}
and (\ref{BM mody ciagle}), (\ref{BM mody ciagle 2}) we can obtain 
formula for energy-momentum tensor evaluated at free vacuum\footnote{Subscript $b$ denotes bare quantities which need regularization. For a time being we introduce an implicit cutoff $\Lambda$ both for integrals and sums which we will taken to infinity at the end.}
\begin{equation}
    \bra{\Theta}\hat{T}_{\mu\nu}\ket{\Theta}_b=
    \frac{1}{a^2\chi^2}\int_0^{+\infty}k\dd k \left( g_{\mu\nu}+2\delta_{\mu}^{\chi}\delta_{\nu}^{\chi} \right) \label{BM too1}
\end{equation}
and using (\ref{BM mody dyskretne}), (\ref{BM mody dyskretne 2}) and (\ref{BM ep tensor}) we can obtain formula for energy-momentum tensor evaluated at Dirichlet vacuum inside cavity 
\begin{equation}
    \bra{0}\hat{T}_{\mu\nu}\ket{0}_b=\sum_{n=1}^{\infty}\frac{n\pi}{2 B^2 \log^2 \left(\frac{B}{A}\right)}\left( g_{\mu\nu}+2\delta_{\mu}^{\chi}\delta_{\nu}^{\chi} \right).\label{BM too2}
\end{equation}
Using expressions  (\ref{BM too1}), (\ref{BM too2}) and (\ref{BM energy}) we can calculate bare Casimir energy 
\begin{equation}
    E_{bc}=\int_A^B\frac{1}{a\chi}\sum_{n=1}^{\infty}\frac{a^2n\pi}{2 \log^2 \left(\frac{B}{A}\right)}\dd\chi- \int_A^B\frac{1}{a\chi}\int_0^{+\infty}k\dd k\dd\chi,
\end{equation}
after integrating over space we get
\begin{equation}
    E_{bc}=\frac{\log\left(\frac{B}{A}\right)}{a}
    \left(\sum_{n=1}^{\infty}\frac{a^2n\pi}{2 \log^2 \left(\frac{B}{A}\right)}- \int_0^{+\infty}k\dd k\right).\label{BM -1}
\end{equation}
The sum and integral in the expression (\ref{BM -1}) are divergent, therefore regularization is necessary. A substitution $k= n \sqrt{\frac{a^2\pi}{2\log\frac{B}{A}}}$ yields
\begin{equation}
    E_{bc}=  \frac{a \pi }{2 \log\frac{B}{A}}
    \left(\sum_{n=0}^{\infty} n- \int_0^{+\infty}n\dd n\right). \label{BM tem}
\end{equation}
The expression in this form is still ill-defined, as a difference of two infinities. However, we can introduce a cutt-off $\Lambda$ in both the sum and the integral.
With the expression in this form, we can use the Euler-Maclaurin summation formula and finally obtain by taking $\Lambda\rightarrow \infty$ a regularized expression

\begin{equation}
    E_c=  -\frac{a \pi }{24 \log\frac{B}{A}}.\label{BM -2}
\end{equation}
Few comments are in place. First of all, we defined Casimir energy as the difference between energies in the cavity's volume of a free theory and of the field with Dirichlet boundary conditions. In full generality, it should be defined as follows: an energy of the field with the imposed boundary condition from both plates  minus the sum of energies with only one boundary plus the one without any boundaries. It reads
\begin{equation}
E_c=E_{total}-E_{A}-E_{B}+E_{vacuum}.
\end{equation}
Let us notice that in this way energies outisde of cavity cancel formally. For the geometry considered here, these definitions are equivalent (as adding a single mirror does not change energy in the Rindler wedge \cite{Davies:1976hi}). Physical reason for this simplification is the lack of self-force (i.e. a force acting on a single mirror in a vacuum). In general case, our calculation would give only an addition coming from the presence of the second mirror (see \cite{Kenneth:2006vr, Emig:2007cf, Fermi:2015xua}).
One should also notice that the same result could be obtained if we have defined energy as a appropriately regularized (by zeta-regularization) sum over all frequencies $\omega_n =\frac{an\pi}{2\log\frac{B}{A}}$. \\
Equation (\ref{BM -2}) leads to the attractive Casimir force acting on plate $B$
\begin{equation}
    F_c^I=-\frac{\partial E_c}{\partial B}=-\frac{a\pi}{24B\log^2\left(\frac{B}{A}\right)} \label{BM zla sila}
\end{equation}
and analogously acting on A is the force 
\begin{equation}
    F_c^I=-\frac{\partial E_c}{\partial A}=\frac{a\pi}{24A\log^2\left(\frac{B}{A}\right)}
\end{equation}
One immediately notices that those results are $a$-dependent. It should be no surprise since the vector $\partial_\tau$ depends upon this parameter. \\ 
One could be tempted to define alternatively the force as a field pressure associated to the spatial Killing vector. However, no vector field generates a symmetry of the Rindler space (in particular $\partial_x$ does not preserve the horizon) and thus we are not equipped with a Noether formula analogous to \eqref{BM energy}.

\subsection{Casimir force as a field pressure \label{BM roz B}}
On the other hand, an inertial observer is equipped with a pressure formula associated with $\partial_x$ so we propose an alternative approach.
We can identify the Casimir force that the inertial observer sees when the cavity rests in its frame with the force by point-like dynamometers attached to the plates.
The construction of the Rindler coordinates indicates that 
these quantities
can justifiably be considered equal \cite{Rindler1969}. 
Our considerations will take place in the inertial frame in which at  $ t = 0 $ the plates are at rest (Fig. \ref{BM PIC}).\\
From the inertial observer point of view, a field pressure acting on the right plate can be written as the energy momentum tensor contracted with the spatial Killing vector $ N_M^{\nu} \partial_\nu= \partial_x $ and a unit vector normal to the surface of a plate $ v^{\mu} \partial_\mu = \partial_{\chi} $ \cite{Wald:1984rg}.

The pressure acting on the right plate from the cavity site can be written as\footnote{We need to compute pressure exactly at the plate. Otherwise, the result is influenced by a change of the momentum of electromagnetic field in the surrounded volume. The momentum density is no longer constant in the accelerating case. This may lead to problems in $4$ dimensions, as the stress energy at the boundary might be divergent.}
\begin{equation}
p_I=\bra{0}\hat{T}_{\mu\nu}v^{\mu}N_M^{\nu}\ket{0}, \label{BM PI}
\end{equation} 
and the pressure acting on the same plate from the empty space site is
\begin{equation}
p_O=-\bra{\Theta}\hat{T}_{\mu\nu}v^{\mu}N_M^{\nu}\ket{\Theta}. \label{BM PO}
\end{equation}
On the other hand, if we consider a case of a single mirror than the corresponding pressures are
\begin{equation}
p_I^B,\quad p_O^B=p_O
\end{equation}
where we used the fact that in the both system with and without mirror placed in $A$ the vacuum on the right from mirror $B$ is the same.

To write Casimir force we need to express the Killing spatial vector in Rindler coordinates.
 \begin{equation}
     N_M^{\nu}\partial_\nu=\partial_x=\frac{\partial x}{\partial \chi}\partial_{\chi}+\frac{\partial x}{\partial\tau}\partial_{\tau}=a\chi \sinh a\tau \partial_{\tau}+\cosh a\tau \partial_{\chi}.
 \end{equation}
At the time of $\tau = 0 $, the normalized vector normal to the surface of the plate coincides with the Killing spatial vector, which indicates that in the selected coordinate system at the time of $ \tau = 0 $ the plates rest.

Casimir force can be written as
\begin{align}
    F_c=p_I+p_O^B&=a\chi \sinh a\tau \bra{0}\hat{T}_{\chi\tau}\ket{0} + \cosh a\tau \bra{0}\hat{T}_{\chi\chi}\ket{0}
    \nonumber
    \\
   &-a\chi \sinh a\tau \bra{\Theta}\hat{T}_{\chi\tau}\ket{\Theta}  - \cosh a\tau \bra{\Theta}\hat{T}_{\chi\chi}\ket{\Theta} \label{BM Fp}
\end{align}

To determine the force, we need to use four expected values of the energy momentum tensor on the vacuum states obtained in (\ref{BM too1}) and (\ref{BM too2}) . We will evaluate each of them in $ \chi = B $. To this end we will use mode decomposition introduced earlier. Calculation for the left mirror is analogous. The only difference is that in this case pressure $p_I$ is negative and $p_O$ is positive. In fact, one should regularize $p_I$ and $p_0$ separately and take their sum at the end. However, as it was already discussed, in this particular case (Rindler vacuum), there is no self-interaction of a single mirror, it is the same as regularizing the whole sum. We discuss generalizations in the Sec. \ref{MK roz D}.
Altogether we obtain
\begin{equation}
    F_{bc}= \frac{\cosh a\tau}{a^2B^2}  \left(  \sum_{n=1}^{\infty}\frac{n \pi a^2}{2 \log^2\frac{B}{A}} -  \int_0^{\infty}k\dd k \right).
\end{equation}
The last step, as in the case of (\ref{BM tem}) is regularization. We use a substitution $k\rightarrow\sqrt{\frac{n\pi a^2}{2\log^2\frac{B}{A}}}$ and Euler-Maclaurin summation formula to finally get
\begin{equation}
    F_c=- \frac{\pi \cosh{a\tau}}{24B^2\log^2\frac{B}{A}}. \label{BM 2}
\end{equation}
Let us note that the obtained force depends on time.
However, from the beginning of the derivation, we knew that the final force will be equivalent to the Casimir force acting on the plate only at the moment of $ \tau = 0 $. Then Casimir force will simplify itself to
\begin{equation}
    F_c^{II}=- \frac{\pi }{24B^2\log^2\frac{B}{A}}. \label{BM 22}
\end{equation}
The presence of the term $\cosh{a\tau}$ in (\ref{BM 2}) can easily be explained by examining how the twoforce \cite{Rindler1969} transforms under a boost from a plate $B$ rest frame to a frame in which the plate is moving with the instantaneous velocity $v$.
\begin{equation}
f=\begin{bmatrix}
0 \\
- \frac{\pi }{24B^2\log^2\frac{B}{A}} 
\end{bmatrix}
\longrightarrow 
\begin{bmatrix}
- \frac{\gamma(v)v\pi }{24B^2\log^2\frac{B}{A}} \\
- \frac{\gamma(v)\pi }{24B^2\log^2\frac{B}{A}} 
\end{bmatrix}.
\end{equation}
Knowing the trajectory of plate $B$, we can express $v$ and $\gamma(v)$ in terms of $\tau$

\begin{equation}
f=\begin{bmatrix}
- \frac{\pi\sinh a\tau  }{24B^2\log^2\frac{B}{A}}  \\
- \frac{\pi\cosh a\tau  }{24B^2\log^2\frac{B}{A}} 
\end{bmatrix}.
\label{4force}
\end{equation}
This derivation shows that time dependence of the force given by \eqref{BM 2} could be understood as the effect of a simple Lorentz transformation from the cavity rest frame. We showed that spacelike component of the twoforce \eqref{4force} is exactly equal \eqref{BM 2}. Additionally, a timelike component of the twoforce \eqref{4force} represents energy transfer from the field to the plate.  
It means that more effort have to be put to maintain uniform acceleration of plate $B$ due to Casimir effect. Analogical calculations are showing that acceleration of plate $A$ requires less effort.

\subsection{Comparison \label{BM roz C}}
Expressions (\ref{BM zla sila}) and (\ref{BM 22}) obtained in both cases are different! 
It means that at least one definition is based on  a non-physical assumption.
We should feel quite safe about equation (\ref{BM 22}) because its canonical force which follows from Noether theorem and Hamiltonian formalism \cite{Wald:1984rg}.

On the other hand significant suspicion should fall on the expression (\ref{BM zla sila}) determined by energy considerations. The calculated force depends on $a$, i.e. the acceleration on the reference hyperbola on which the Rindler's time is measured. $a$ is a transformation parameter that can take any value and indication of the dynamometer attached to the plate should not depend on it. Therefore, it should be obvious that the expression (\ref{BM zla sila}) in its present form does not make any sense.
However, it is possible to fix $F_c^I$ obtained in (\ref{BM zla sila}) by noting that in Rindler's frame at different points time run differently. We define the coordinate time on the reference hyperbola $\tau_0$ and the local time on the hyperbola on which the right plate lies. The derivation 
of (\ref{BM zla sila})
took into account only coordinate quantities. If we translate them into a language of local quantities we get
\begin{equation}
    F_c^I=\frac{\dd p}{\dd \tau_0}=\frac{\dd\tau}{\dd\tau_0}\frac{\dd p}{\dd \tau}=\frac{1}{aB}\frac{\dd p}{\dd\tau}=\frac{1}{aB}F_c^{II}.
\end{equation}
This means that after taking into account the differences between the coordinate time and the local time in the $F_c^I$, We will obtain a match between $F_c^{I}$ obtained in (\ref{BM zla sila}) and $F_c^{II}$ obtained in (\ref{BM 22}).  \\
We should expect that when plates are moved far from the Rindler horizon $(A,B\rightarrow\infty)$ (it means taking their accelerations to zero), while keeping fixed distance between them $(L=B-A=$const.$)$ Casimir force will reduce to the usual stationary expression. Indeed
\begin{equation}
    F_c^{II}=-\frac{\pi}{24B^2\log^2\left(\frac{B}{A}\right)}\rightarrow -\frac{\pi}{24(B-A)^2}=-\frac{\pi}{24L^2}
\end{equation}
which confirms our expectations and shows that force $F_c^{II}$ and "repaired" force $F_c^{I}$ behave correctly in the zero acceleration limit. \\
The result \ref{BM roz A} can be repaired in such a simple way only because we consider flat plates. When we change the geometry of the system, this method will become impossible to use. Approach \ref{BM roz B} will give the correct results whenever we are able to introduce a comoving frame, i.e. when the mirrors move along any trajectory in a flat space or in a conformally  flat spacetime. Below we present an alternative method much more computationally effective that will work in all cases in which the method of a comoving frame works - \ref{BM roz B}.

The above analysis indicates that operating in non-inertial reference frames requires awareness about many non-intuitive subtleties such as inability to synchronize the passage of time at different points in space, unclear relation between Casimir force and Casimir energy or different acceleration in different space points.
Neglecting these subtleties lead to improper formulas for Casimir Force, as can be seen in
 \cite{Zhao2011, Saharian2004, Wilson2019, Lambiase2016, Grigoryan2016, Assel2015, Esposito2008, Roman1986, Caldwell2002, Buoninfante2018, Bezerra2014, Ford1997, Sorge2005, Zhang2017}.

\subsection{The question of the state outside \label{MK roz D}}

So far we implicitly assumed that the state outside our cavity is Rindler vacuum. 
However, Rindler vacuum behaves singularly at the horizon, thus it is not physically viable state. In fact, we should take some Hadamard state instead\footnote{We leave open question about the proper definition of Hadamard property for systems with boundaries, assuming only that the property is satisfied in the bulk.}. From the very construction of the quantum energy--momentum tensor operator $\hat{T}_{\mu \nu}$ \cite{Wald:1977up} it follows that the difference would be finite and thus one cannot distinguish between those possibilities on purely theoretical grounds. In fact, many such possibilities could be realized experimentally. Indeed, let us at first imagine an accelerating rocket such that an observer inside does not detect any particles -- it means filled with the static vacuum. Aforementioned observer could install a small (in comparison with the rocket length) and stationary cavity and measure Casimir force in such a setting. Then, they would find result close to the one described in the previous paragraphs. \\
On the other hand, one could start with an inertial laboratory (say, in outer space) in which only cavity is moving. Then, one would expect to find a little bit different results -- we relegates details to the next section because the answer is easier to find using conformal methods. However, in any case the difference is related to the change of the force acting on the single accelerating mirror and not to the additional force related to the presence of the cavity.\\
In any analogous experiment one should be very careful in distinguishing between those two effects.

\section{Energy--momentum tensor from conformal anomaly} 
Calculations presented in the previous sections are straightforward but a little bit laborious. If one were tempted to find the force acting upon a mirror in a different situation (e.g. for different trajectories or on some non-trivial background), he or she would have to repeat it all over again: find modes, integrate them together to obtain $T_{\mu \nu}$, regulate their result in a suitable manner... In this section we show how this can be avoided by the usage of the conformal symmetry of a (classical) theory at hand. In particular, both massless Klein--Gordon field in $1+1$ dimensions and Maxwell field in $3+1$ dimensions posses this symmetry.
\subsection{Conformal anomaly \label{MK roz III}}
It is well--known fact that under conformal transformation 
\begin{align}
    \begin{split}
        g_{\mu \nu}(x) &\mapsto e^{2 \sigma(x)} g_{\mu \nu}(x) \\
        \phi(x) &\mapsto e^{-n\sigma(x)} \phi(x)
    \end{split}
\end{align}
classical energy--momentum tensor of a conformal field theory transforms covariantly: $T_{\mu \nu} \mapsto e^{-2s \sigma} T_{\mu \nu}$ (constant $s$ is here dimension dependent and $n$ depends upon dimension and also spin). Unfortunately, it is no longer true at the quantum level. This can be traced back to the conformal anomaly, namely to the fact that $\langle \hat{T}^\mu_{\ \mu} \rangle \neq 0$. Interestingly enough, that trace has a universal, geometric structure -- it depends only upon a curvature of a background and not on a state on which it is evaluated. In two dimensions it is given by \cite{Christensen:1978gi}:
\begin{equation}
    \langle \hat{T}^\mu_{\ \mu} \rangle_{d=2} = - \frac{R}{24 \pi}
\end{equation}
whereas in four dimensions:
\begin{equation}
    \langle \hat{T}^\mu_{\ \mu} \rangle_{d=4} = \frac{1}{2880\pi^2} \left[a C_{\lambda \mu \nu \xi}C^{\lambda \mu \nu \xi} + b \left(R_{\mu \nu} R^{\mu \nu} - \frac{1}{3}R^2 \right) + c R_{;\mu}^{\ \ ;\mu} \right],
\end{equation}
where $a=b=-1=-c$ for the conformally coupled massless scalar field and $a=13$, $b=-62$, $c=-18$ for the Maxwell field. \\Surprisingly, it is enough to know merely conformal anomaly to recover transformation law for $\langle \hat{T}^\mu_{\ \nu} \rangle$ upon conformal transformations. Although it was derived in the full generality in small dimensions of our interest, the result is lengthy. For all practical purposes, we can assume that our metric is of the form $g_{\mu \nu} = e^{2 \sigma} \eta_{\mu \nu}$. Then, we have \cite{Brown:1977sj}:
\begin{align}
    \begin{split}
        \langle \hat{T}_{\mu \nu} \rangle_{d=2} &=  \langle \hat{T}_{\mu \nu} \rangle_0 + \frac{1}{12\pi} \left[\sigma_{;\nu \mu} -\sigma_{;\mu} \sigma_{;\nu}+ \eta_{\mu \nu} \left( \frac{1}{2} \sigma_{;\xi} \sigma^{;\xi} - \sigma^{;\xi}_{\ \ ;\xi} \right) \right] \\
                \langle \hat{T}_{\mu \nu} \rangle_{d=4} &= e^{-2 \sigma} \langle \hat{T}_{\mu \nu} \rangle_0 - \frac{1}{16\pi^2} \left( \frac{c}{1080} ^{(1)}H_{\mu \nu} + \frac{b}{180} ^{(3)}H_{\mu \nu} \right),  \label{transformacja}
    \end{split}
\end{align}
where $\langle \hat{T}_{\mu \nu} \rangle_0$ is expectation value of energy--momentum tensor of a state pull-backed by conformal transformation to the Minkowski spacetime, all curvature tensors are evaluated with respect to the metric $g_{ab}$ and
\begin{align}
    \begin{split}
        ^{(1)}H_{\mu \nu} &= 2R_{;\mu \nu} - 2g_{\mu \nu} R_{;\lambda}^{\ \ ;\lambda} - \frac{1}{2} g_{\mu \nu} R^2 + 2RR_{\mu \nu}, \\
        ^{(3)}H_{\mu \nu} &= R_{\mu}^{\ \ \rho} R_{\rho \nu} - \frac{2}{3} R R_{\mu \nu} - \frac{1}{2} R_{\rho \lambda} R^{\rho \lambda} g_{\mu \nu} + \frac{1}{4}R^2 g_{\mu \nu}.
    \end{split}
\end{align}
Equations (19) and (42) in \cite{Brown:1977sj} do not contain homogeneous terms $e^{-2 \sigma} \langle \hat{T}_{\mu \nu} \rangle_0$ because it was assumed that they vanishes, it means that the state in question was a conformal vacuum. One can easily convince 
oneself that for a general state, the only change is in the presence of that homogeneous part.  \newline
Equipped with that knowledge, we can lay down our strategy. We consider a set of possibly moving mirrors in a possibly curved (but conformally flat) spacetime. We look for a coordinate system $(t, x^i)$ and a conformal factor $e^{\sigma}$ such that
\begin{equation}
    g = e^{2 \sigma} \left(\dd t^2 - \delta_{ij} \dd x^i \dd x^j \right)
\end{equation}
and mirrors trajectories are given by $x = \textrm{const.}$ 
\footnote{In general, we could even consider mirrors which are not infinite planes. Then, they would be located at the locus of equation $f(x^i) = 0$ for some functions $f$.}
Then, we can calculate $\langle T_{ab} \rangle_0$ for stationary vacuum with appropriate boundary conditions and transform it back to our background using \eqref{transformacja}. We present examples of that strategy in the next section.
\subsection{Allowed boundary conditions}
Before that, let us comment on the question what kinds of mirrors (it means, what kinds of boundary conditions) can be investigated by the means of conformal method. Obviously, homogeneous Dirichlet boundary conditions $\phi(t,x) = 0$ are invariant upon rescaling by a non-vanishing factor. However, any non-homonogenity would destroy that property. Also, the Neumann boundary condition $n^a \phi_{,a} = 0$ with $n^a$ being spacelike vector normal to the surface is not conformally invariant -- it would be transformed into a point- and scale factor-dependent Robin boundary condition. However, if scale factor is constant along the direction perpendicular to the boundary (as it is the case in any cosmological scenario, where it depends only upon time), also Neumann condition is preserved.  \\
Usually one considers electromagnetic Casimir effect. 
Then, mirrors are perfect conductors so that the magnetic field is tangent and the electric field normal to the boundary.
This can be formulated as a condition $F_{\mu \nu} t^\mu s^\nu = 0$, for any vectors $t^\mu, s^\nu$ tangent to the boundary (either spacelike or timelike).
Since the field strength tensor $F_{\mu \nu}$ is conformally invariant, also this boundary condition is preserved. 
Thus, we see that despite its limitations, method proposed here covers a broad spectrum of physically interesting configurations. \\
Another question one should ask themselves is whether state living inside the cavity really corresponds to the vacuum state in Minkowski spacetime. At least in two circumstances the answer should be affirmative. If $\sigma$ goes to $0$ fast enough as $t$ goes to minus infinity, then our situation was Minowskian at first and thus should have the same vacuum in the asymptotic past. On the other hand, if $\partial_t$ is a Killing vector, then we have well-defined notion of a static vacuum both for $g$ and $\eta$ and they are going to be again connected by a conformal transformation. We will see examples of both types in the next section.
\section{Casimir effect on a conformally flat background}
\subsection{Rindler mirrors \label{MK roz IVA}}
Let us start by re-examing accelerating mirrors considered in the previous sections.
In the Rindler coordinates $(\tau, \chi)$, mirrors are located at $\chi = A, B = const.$ and Minkowski metric takes the form:
\begin{equation}
    \dd s^2 = (a \chi)^2 \dd\tau^2 - \dd\chi^2.
\end{equation}
Let us introduce a new coordinate $\xi$ such that $\chi = \frac{1}{a} e^{a \xi}$. 
Metric now is explicitly conformally flat:
\begin{equation}
    \dd s^2 =  e^{2a\xi} \left( \dd\tau^2 - \dd\xi^2 \right)
\end{equation}
so we can use \eqref{transformacja} with $\sigma = a \xi$. As discussed before, conformal transfomation maps the Minkowski vacuum into the Rindler vacuum and preserves boundary conditions, so we have:
\begin{equation}
    \langle \hat{T}_{\mu \nu} \rangle_{Rindler} = \langle \hat{T}_{\mu \nu} \rangle_{Mink} - \frac{a^2}{12\pi} \left[ \delta^\xi_{\ \mu} \delta^\xi_{\ \nu} + \frac{1}{2} \eta_{\mu \nu}  \right]. \label{conf_ridn} 
\end{equation}
Substituting \cite{Birrell:1982ix}
\begin{equation}
    \langle \hat{T}_{\mu \nu} \rangle_{Mink} = - \frac{\pi}{24(\xi_B - \xi_A)^2} \left(\eta_{\mu \nu} + 2 \delta_\mu^\xi \delta_\nu^\xi \right) = - \frac{a^2\pi}{24\log^2\left(\frac{B}{A} \right)} \left(\eta_{\mu \nu} + 2 \delta_\mu^\xi \delta_\nu^\xi \right)
\end{equation}
we obtain energy--momentum tensor for accelerating mirrors:
\begin{equation}
    \langle \hat{T}_{\mu \nu} \rangle_{Rindler} dx^\mu dx^\nu = \frac{a^2\pi}{24\log^2\left(\frac{B}{A} \right)} \left(d\tau^2 + \frac{1}{a^2 \chi^2}d \chi^2 \right) - \frac{a^2}{24\pi} \left(d\tau^2 + \frac{1}{a^2 \chi^2}d\chi^2 \right)
\end{equation}
The second term represents energy-momentum of the Rindler vacuum and thus must be subtracted before calculating the force. Contracting with vectors $\partial_{\chi}$ and $\partial_x$ we finally obtain our desired result:
\begin{equation}
    F = - \frac{\pi \cosh a \tau}{24 B^2 \log^2 \frac{B}{A}}
\end{equation}
exactly as before. \\
 Inclusion of the second term leads to the additional pressure which is independent from the cavity size. It should be taken into account when accelerating cavity is sorrounded rather by the Minkowski not Rindler vacuum.  \\
One could also consider one mirror which moves freely (but with subliminal velocities) in $1+1$ Minkowski spacetime.
This problem was solved in \cite{Davies:1976hi} and as long as acceleration is not uniform, mirror emitted radiation (as was seen in the form of the energy--momentum tensor).
Although authors used mode decomposition in their calculations, they conjectured some (unclear at that point) connection between that radiation and the conformal anomaly (despite the fact that $\langle \hat{T}^{\mu}_{\ \mu} \rangle = 0$, since the background is flat).
One can check that \eqref{transformacja} allows to reproduce their results and thus this work provides an explanation for that connection.
\subsection{Mirrors in an expanding universe \label{MK roz IVB}}
Let us consider the general FLRW background with $k=0$:
\begin{equation}
    \dd s^2 = a^2(\eta) \left(\dd\eta^2 - \delta_{ij} \dd x^i \dd x^j \right)
\end{equation}
which is explicitly conformally flat with $\sigma = \log a$. $\eta$ is a conformal (in contrast with a cosmological) time. We put two infinite, flat conductors at $z = z_{1/2}$. As discussed before, if $a(\eta)$ approaches $1$ fast enough when $\eta \rightarrow -\infty$, we can expect the state within this cavity to be equivalent to the Minkowski vacuum for Maxwell field bounded by two conductors (it means by the usual state of the original Casimir effect). Let us calculate useful curvature tensors:
\begin{align}
    \begin{split}
        R_{\mu \nu}\dd x^\mu \dd x^\nu &= \frac{3 \left((a')^2 - aa'' \right)}{a^2} \dd\eta^2 + \delta_{ij} \frac{(a')^2 + aa''}{a^2} \dd x^i \dd x^j \\
        R &= - \frac{6a''}{a^3} \\
        ^{(1)}H_{\mu \nu} \dd x^\mu \dd x^\nu &= 6 a^{-5} (3 (-8(a')^2 a'' + a (a'')^2 + 2 a a' a''') \dd\eta^2\\
        &+\delta_{ij} (-20(a')^2 a'' +7a(a'')^2 + 10a a' a''' - 2a^2 a'''') \dd x^i \dd x^j)  \\
        ^{(3)}H_{\mu \nu}\dd x^\mu \dd x^\nu &= \frac{3(a')^4}{a^6} \dd\eta^2 + \delta_{ij} \frac{5(a')^4 - 4a(a')^2 a''}{a^6}\dd x^i \dd x^j
    \end{split}
\end{align}
As an explicit example let us take the de Sitter spacetime in which $a = \frac{1}{H\eta}$ where $H$ is the Hubble parameter. This gives
\begin{equation}
    \langle\hat{T}_{\mu \nu} \rangle = H^2 \eta^2 \langle\hat{T}_{\mu \nu} \rangle_0 - \frac{H^4 b}{960\pi^2} g_{\mu \nu}.
\end{equation}
Not surprisingly, we see that the non-homogeneous term becomes important only for distances between mirrors of order $\mathcal{O}(H^{-1}c)$  and the conformal homogeneous correction $H^2 \eta^2$ matters only for times of the order of a lifetime of a universe. Moreover, local inhomogeneity would surpass this effect. It is thus way beyond the possibility of measurement. However, it is not so obvious if one considers analogue experiments. As in the Rindler case, the second term depends heavily upon the state on the other part of the mirror, in particular it would have if there was a conformal vacuum outside.\\ 
Pressure of the Casimir force itself reads
\begin{equation}
    F = H^2 \eta^2 F_{Mink} + \frac{H^4 b}{960\pi^2},
\end{equation}
so (not surprisingly) the gravitational effect is very small. It depends on the possibility of keeping two mirrors comoving with the expanding universe with  high accuracy and it will be invisible among effects created by local curvature. However, one can hope to observe an analogous effect for example in
Bose--Einstein condensate experimentor in metamaterial analog experiment.

\section{Conclusions}
We used the method of Casimir energy to show how that the Casimir force in the accelerated frame of reference is incompatible with the analogous result obtained by means of evaluating 
the pressure acting on the plate. We have discussed the correctness of these incompatible results to conclude that the method known from the original Casimir paper couldn't be easily generalized to the case of more complicated geometries. In section II D, we also showed how to fix the incorrect result obtained by the Casimir energy in case of mirrors with very simple geometry. The results received in this way have allowed us to understand possible difficulties in the generalization of the Casimir force to the more complicated space-times. In Sec. III A, we have developed an effective scheme of finding Casimir force in the case of conformally flat space-times. We have used a conformal anomaly to show the relation between properties of the transformation of the momentum energy tensor and the Casimir force. We applied this approach to the case of uniformly accelerating plates in 1+1 dimensional space-time in Sec. IV A. And finally, in Sec. IV B we also derived Casimir force for plates in 3+1 dimensional de Sitter spacetime.
\begin{acknowledgments}
MK was financed from budgetary funds for science in 2018-2022 as a research project under the program "Diamentowy Grant".
\end{acknowledgments}

\end{document}